\documentclass[journal]{IEEEtran}
\usepackage{cite}
\usepackage{amsmath,amssymb,amsfonts}
\usepackage{algorithmic}
\usepackage{graphicx}
\usepackage{textcomp}
\usepackage[inline]{enumitem}
\usepackage{balance}
\usepackage{booktabs}
\usepackage{multirow}
\usepackage{bm}
\usepackage{physics}
\usepackage{unit}
\usepackage{url}
\usepackage{threeparttable}
\usepackage[super]{nth}
\makeatletter
\let\MYcaption\@makecaption
\makeatother
\usepackage[font=footnotesize]{subcaption}
\makeatletter
\let\@makecaption\MYcaption
\makeatother
\usepackage{caption}
\def\BibTeX{{\rm B\kern-.05em{\sc i\kern-.025em b}\kern-.08em
    T\kern-.1667em\lower.7ex\hbox{E}\kern-.125emX}}
\begin{document}

\title{Radar-Based Identification of Individuals Using Heartbeat Features Extracted from Signal Amplitude and Phase}

\author{Haruto~Kobayashi and Takuya~Sakamoto
\thanks{H.~Kobayashi and T.~Sakamoto are with the Department of Electrical Engineering, Graduate School of Engineering, Kyoto University, Kyoto 615-8510, Japan.}
\thanks{H.~Kobayashi is also with Hitachi, Ltd., 1-280, Higashi-Koigakubo, Kokubunjishi, Tokyo 185-8601, Japan.}
}
\markboth{}
{Kobayashi and Sakamoto: Radar-Based Identification of Individuals Using Heartbeat Features}

\maketitle

\begin{abstract}
This study proposes a non-contact method for identifying individuals through the use of heartbeat features measured with millimeter-wave radar. Although complex-valued radar signal spectrograms are commonly used for this task, little attention has been paid to the choice of signal components, namely, whether to use amplitude, phase, or the complex signal itself. Although spectrograms can be constructed independently from amplitude or phase information, their respective contributions to identification accuracy remain unclear. To address this issue, we first evaluate identification performance using spectrograms derived separately from amplitude, phase, and complex signals. We then propose a feature fusion method that integrates these three representations to enhance identification accuracy. Experiments conducted with a 79-GHz radar system and involving six participants achieved an identification accuracy of 97.67\%, demonstrating the effectiveness of the proposed component-wise analysis and integration approach.
\end{abstract}

\begin{IEEEkeywords}
 Classification, heartbeat features, individual identification, millimeter-wave radar, non-contact sensing
\end{IEEEkeywords}

\IEEEpeerreviewmaketitle

\section{Introduction}
Physiological signal monitoring has traditionally relied on contact-based sensors, which can be uncomfortable or inconvenient for continuous use. Consequently, non-contact measurement methods have attracted increasing attention, particularly in healthcare applications. Among them, radar technology stands out for its ability to detect vital signs such as respiration~\cite{Koda, https://doi.org/10.1109/TMTT.2019.2933199} and heartbeat~\cite{https://doi.org/10.1109/TIM.2023.3300471, https://doi.org/10.1109/LSENS.2023.3322287} without physically constraining the user. As a result, radar-based non-contact sensing has become an active research area in recent years~\cite{https://doi.org/10.1109/JPROC.2023.3244362, https://doi.org/10.1109/MAES.2019.2933971}.

Beyond health monitoring, radar-based measurement of physiological signals also exhibits potential for identifying individuals ~\cite{https://doi.org/10.1016/j.compeleceng.2024.109485}, and several radar-based identification techniques exploiting features extracted from respiration and heartbeat signals have been proposed ~\cite{https://doi.org/10.1145/3117811.3117839, https://doi.org/10.1109/JETCAS.2018.2818181, https://doi.org/10.1016/j.bspc.2021.103306, https://doi.org/10.1109/ACCESS.2023.3328641, https://doi.org/10.1109/JSEN.2024.3353256, https://doi.org/10.1109/JIOT.2024.3358548, https://doi.org/10.1109/ACCESS.2024.3478814, Kobayashi2024}. We previously proposed a radar-based individual identification method that combined respiratory and heartbeat characteristics and achieved an identification accuracy of 96.7\% on data from six participants~\cite{Kobayashi2024}. Building on this result, the present study aims to further improve identification performance by analyzing the contributions of different signal components.

In respect to identification of individuals, heartbeat signals offer a notable advantage over respiratory signals: they occur at higher frequencies, enabling feature extraction from much shorter measurement durations. This property allows accurate identification even when only brief data samples are available. For example, in our previous study~\cite{Kobayashi2024}, we achieved identification accuracies of 93.33\% and 85.64\% using only heartbeat features extracted from 60-second and 5-second samples, respectively.

These heartbeat features were extracted by directly processing the complex-valued radar signals. However, it remains unclear which components---amplitude, phase, or the complete complex signal---contribute the most to the accuracy of identification. Since each of these components can be used to generate spectrograms independently, a systematic comparative evaluation is required.

Our evaluation indicates that the amplitude-, phase-, and complex-valued spectrograms each capture distinct aspects of the heartbeat signal, suggesting that they encode complementary information useful for individual identification. Based on this insight, we propose a method that constructs a single composite feature vector by concatenating the features extracted separately from the amplitude, phase, and complex signals. This combined feature vector is then input to a support vector machine (SVM) to perform individual identification.

\section{Heartbeat Feature Extraction}
\subsection{Mel-Frequency Cepstral Coefficients (MFCC) Feature Extraction}
Our previous work proposed a method for extracting heartbeat characteristics for individual identification~\cite{Kobayashi2024}. In this method, MFCCs~\cite{https://doi.org/10.1109/JPROC.2013.2251592} derived from the complex-valued radar signal $s(t)$ are used to represent heartbeat features. We also demonstrated that second-order differentiation enhances the performance of radar-based heartbeat measurement~\cite{itsuki}. Accordingly, the second-order derivative $s''(t) = (\dd[2]/\dd{t}^2)s(t)$ is first computed, then MFCCs are subsequently extracted from $s''(t)$.

First, a short-time Fourier transform (STFT) is applied to $s''(t)$, and the magnitude of the STFT is used as the spectrogram $S(f,t)$ with a rectangular window of width $T_\mathrm{w}$, which is empirically set to $T_\mathrm{w}=2.0$~s. Second, a mel filter bank is applied to the spectrogram $S(t,f)$, which is defined as
\begin{align}
	H_{\ell}(f) & = \begin{cases}
	\displaystyle \frac{2(f - f_{\ell})}{(f_{\ell+1} - f_{\ell})(f_{\ell+2} - f_{\ell})}, & (f_{\ell}\leq f < f_{\ell+1}), \\
	\displaystyle \frac{2(f_{\ell+2} - f)}{(f_{\ell+2} - f_{\ell+1})(f_{\ell+2} - f_{\ell})}, & (f_{\ell+1}\leq f < f_{\ell+2}), \\
	0, & \text{otherwise},
	\end{cases}
\end{align}
for $\ell=0,1,\ldots,L-1$, where $f_{\ell} = \tilde{f}\left\{\exp\left(m_{\ell}/\tilde{m}\right)-1\right\}$ denotes the center frequency of the $\ell$th mel filter corresponding to the mel scale $m_{\ell}$. The mel scale is defined as
\begin{align}
	m_{\ell}= \tilde{m}\frac{\ell}{L+1}\log\left(1 + \frac{f_{\mathrm{s}}}{2\tilde{f}}\right), \label{eq:mel}
\end{align}
for $\ell = 0,1,\ldots,L+1$, where $f_{\mathrm{s}}$ is the sampling frequency of the radar system and $\tilde{m}$ is given by $\tilde{m} = f'/\log\left(f'/\tilde{f} + 1\right)$. In this study, $L=64$, $\tilde{f}=5.0$~Hz, and $f'=1.0$~kHz. Here, $\tilde{f}$ denotes a reference frequency that controls the scaling of the mel-frequency mapping, and is set to $\tilde{f}=5.0$~Hz.

The mel filter bank is applied separately to the positive and negative frequency components as
\begin{align}
	M_{\ell}  & = \int_{0}^{T_0}\int_{0}^{f_\mathrm{s}/2}S(t,f) H_{\ell}(f) \, \dd f\,\dd t, \label{eq:incoherent_int1}   \\
	M_{-\ell} & = \int_{0}^{T_0}\int_{-f_\mathrm{s}/2}^{0}S(t,f) H_{\ell}(-f) \, \dd f\,\dd t, \label{eq:incoherent_int2}
\end{align}
for $0\leq \ell \leq L-1$, where $T_{0}$ denotes the measurement duration. Note that $M_{+0}$ and $M_{-0}$ are treated as distinct variables.

Next, a discrete cosine transform (DCT) is applied to $M_{\ell}$ and $M_{-\ell}$ to obtain the MFCCs $C_{+k}$ and $C_{-k}$ $(k=0,1,\ldots,K-1)$, where the DCT dimension is set to $K=64$. In this study, the Type-II DCT (DCT-II) is employed, and $C_{+0}$ and $C_{-0}$ are regarded as different coefficients.

Finally, from the resulting $2K$-dimensional MFCC vector, lower-order coefficients up to $2K'$ dimensions are retained under the condition $0<K'<K$. The heartbeat feature vector is defined as 
$\bm{r}_{\mathrm{comp}}=[C_{-(K'-1)},C_{-(K'-2)},\ldots,C_{-0},C_{+0},\ldots,C_{K'-1}]^{\mathsf{T}}$,
where $K'=24$ is empirically chosen and $(\cdot)^{\mathsf{T}}$ denotes the transpose operator. The feature vector $\bm{r}_{\mathrm{comp}}\in \mathbb{R}^{2K'\times 1}$ is extracted from the complex-valued radar signal and used for identification of individuals.

\subsection{Amplitude and Phase Feature Extraction}
The feature vector $\bm{r}_\mathrm{comp}$ is directly derived from the complex signal $s(t)$, and therefore contains information related to both the amplitude and phase of $s(t)$. To evaluate the contributions of these components to individual identification, heartbeat features are separately extracted from the second-order derivatives of the amplitude and phase signals.

Here, the amplitude and phase derivatives are defined as $A''(t) = (\dd[2]/\dd{t}^2)|s(t)|$ and $\theta''(t) = (\dd[2]/\dd{t}^2)\,\mathrm{unwrap}\{\angle s(t)\}$, where $\angle s(t)$ denotes the phase of the complex-valued signal $s(t)$, and $\mathrm{unwrap}\{\cdot\}$ denotes the phase unwrapping operator. MFCCs are then calculated from $A''(t)$ and $\theta''(t)$ to obtain the amplitude feature vector $\bm{r}_\mathrm{amp}$ and the phase feature vector $\bm{r}_\mathrm{ph}$, respectively. Since $A''(t)$ and $\theta''(t)$ are real-valued signals, MFCC computation is restricted to the positive-frequency components of their spectra. Consequently, $\bm{r}_\mathrm{amp}$ and $\bm{r}_\mathrm{ph}$ are $K'$-dimensional feature vectors.

\section{Experimental Evaluation}
\subsection{Experimental Setup}
We evaluated the performance of the proposed and conventional features in identification of individuals by performing radar measurement experiments involving six participants. Specifically, individual identification was performed using $\bm{r}_\mathrm{amp}$, $\bm{r}_\mathrm{ph}$, and $\bm{r}_\mathrm{comp}$, as well as their combined feature vector
$\bm{r}_\mathrm{prop}=[\bm{r}_\mathrm{amp}^\mathsf{T}, \bm{r}_\mathrm{ph}^\mathsf{T}, \bm{r}_\mathrm{comp}^\mathsf{T}]^\mathsf{T}$. Table~\ref{tab:methods} summarizes the four methods that used these characteristics.

\begin{table}[tb]
	\caption{Definitions of the Methods used for Individual Identification}
	\centering
	\renewcommand{\arraystretch}{2} 
	\label{tab:methods}
	\begin{tabular}{llccc}
		\toprule
		\textbf{Method}          & \multicolumn{1}{c}{\textbf{Signals}}                                   & \textbf{Feature} & \textbf{Vector}        \\
		\midrule
		Conv. 1                  & $A''(t)=\displaystyle \dv[2]{t}|s(t)|$                                 & Amplitude        & $\bm{r}_\mathrm{amp}$  \\
		Conv. 2                  & $\theta''(t)= \displaystyle \dv[2]{t}\mathrm{unwrap}\qty{\angle s(t)}$ & Phase            & $\bm{r}_\mathrm{ph}$   \\
		Conv. 3 ~\cite{Kobayashi2024} & $s''(t)= \displaystyle \dv[2]{t}s(t)$                                  & Complex signal   & $\bm{r}_\mathrm{comp}$ \\
		Prop.                    & \multicolumn{1}{c}{$A''(t)$ $\&$ $\theta''(t)$ $\&$ $s''(t)$}          & --               & $\bm{r}_\mathrm{prop}$ \\
		\bottomrule
	\end{tabular}
\end{table}

A millimeter-wave array radar system was used in the experimental setup. This was a frequency-modulated continuous-wave (FMCW) radar system with a center frequency of 79.0~GHz, a center wavelength of $\lambda = 3.8$~mm, and a bandwidth of 3.6~GHz. The radar array is composed of a MIMO array with three transmitting and four receiving elements, which can be approximated using a 12-element virtual linear array with element spacings of $\lambda/2$. The slow-time sampling frequency was 100~Hz.

\begin{figure}[tb]
	\centering
	\includegraphics[width=0.7\linewidth]{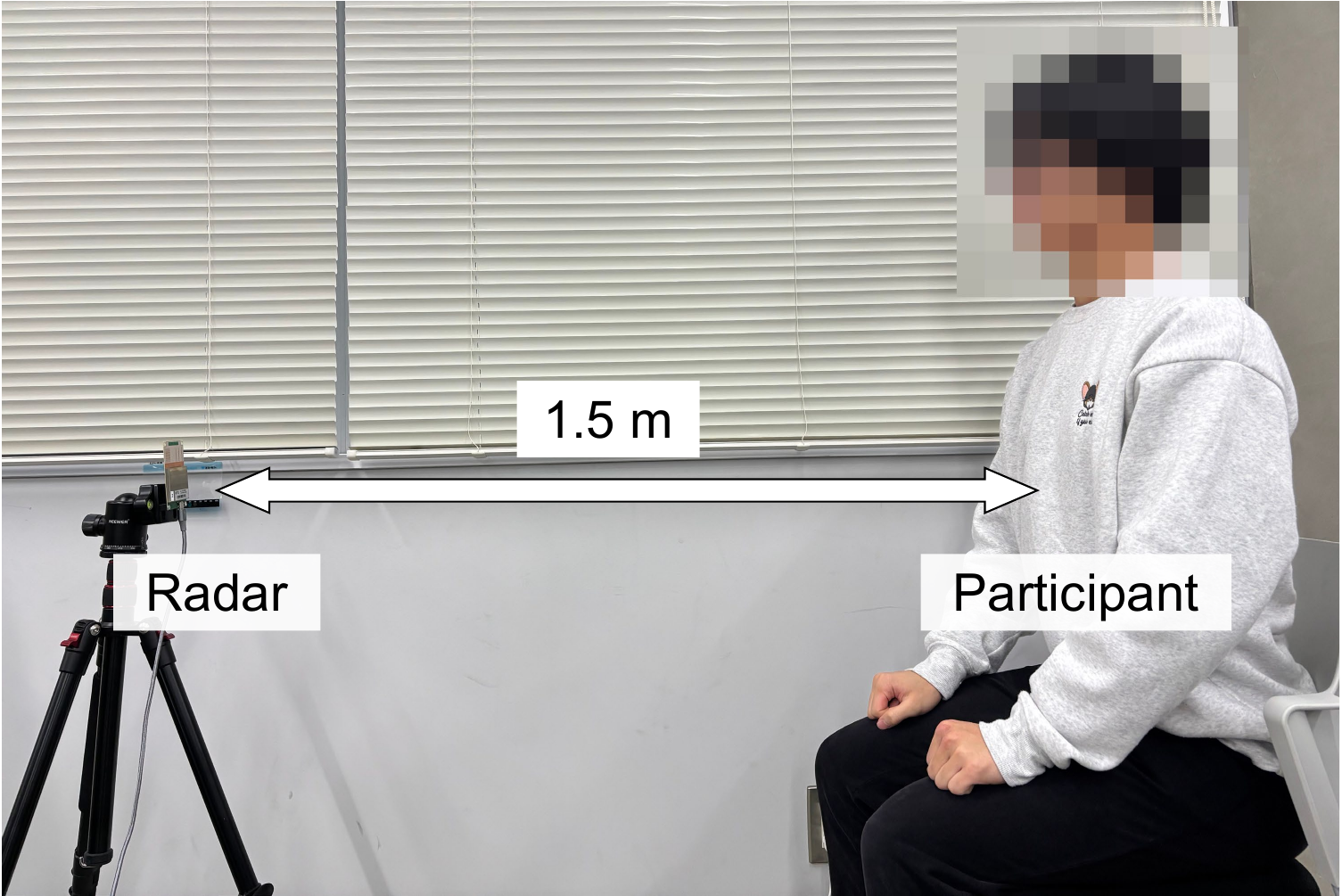}
	\caption{Experimental setup for radar measurement.}
	\label{fig:setup}
\end{figure}

The experiment was repeated on five separate days, with measurements carried out ten times per day, five in the morning and five in the afternoon, with each measurement lasting for 60.0~s. Each participant was seated 1.5~m away from the radar and measured in a resting state. Fig.~\ref{fig:setup} shows the experimental setup. The range and direction of arrival of the target echo are determined using a Fourier transform with respect to the fast time~\cite{FMCW} and the virtual element number based on the beamformer method~\cite{beamformer}. Then, we obtain the radar-received signals $s(t)$ reflected from the target.

Performance in identification of individuals was evaluated using stratified $k$-fold cross-validation with $k=10$, ensuring that each fold preserved the class distribution of the original dataset. In each fold, 270 samples were used for training and 30 samples for validation, with the validation data in each fold being drawn from a single measurement period not included in the training data. For example, in a single fold, data measured during the morning session of the first day were used as validation data while data from all remaining sessions were used for training.

\subsection{Performance Evaluation}
This subsection compares performance in individual identification between data samples of 60~s and 5~s in length. For the 60-s case, a total of 300 samples were used. For the 5-s case, each 60~s recording was divided into 12 segments, resulting in 3,600 samples in total. 

\begin{figure}[tb]
	\centering
	\begin{minipage}{\linewidth}
		\centering
		\includegraphics[width=0.5\linewidth]{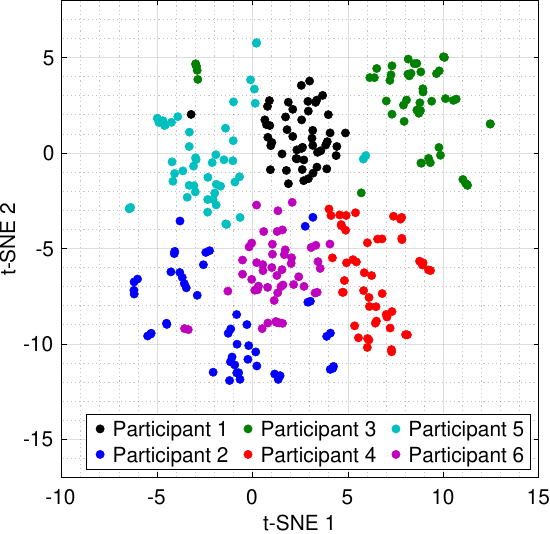}
		\subcaption{}
		\label{fig:tsne_amp}
	\end{minipage}
	\begin{minipage}{\linewidth}
		\centering
		\includegraphics[width=0.5\linewidth]{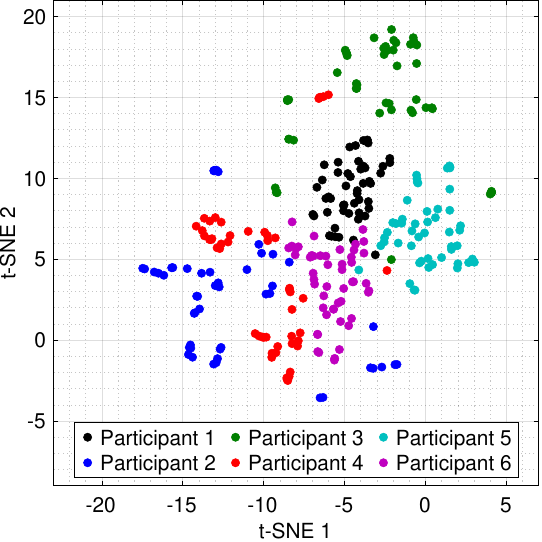}
		\subcaption{}
		\label{fig:tsne_ph}
	\end{minipage}
	\begin{minipage}{\linewidth}
		\centering
		\includegraphics[width=0.5\linewidth]{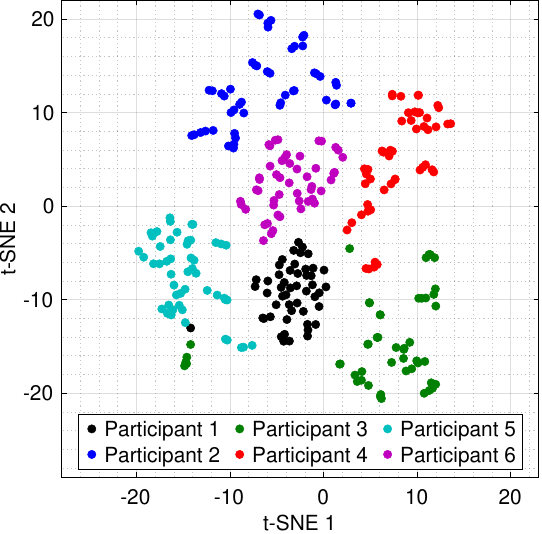}
		\subcaption{}
		\label{fig:tsne_prop}
	\end{minipage}
	\caption{Two-dimensional t-SNE visualizations of the feature vectors (a) $\bm{r}_\mathrm{amp}$, (b) $\bm{r}_\mathrm{ph}$, and (c) $\bm{r}_\mathrm{prop}$ for 60-s data samples.}
	\label{fig:tsne}
\end{figure}

First, for data samples of 60~s, the feature vectors $\bm{r}_\mathrm{amp}$, $\bm{r}_\mathrm{ph}$, and $\bm{r}_\mathrm{prop}$ were visualized using t-distributed stochastic neighbor embedding (t-SNE)~\cite{tSNE}, as shown in Fig.~\ref{fig:tsne}. This visualization indicates that the feature vectors form participant-specific clusters. For $\bm{r}_\mathrm{amp}$ and $\bm{r}_\mathrm{ph}$, it is possible to observe partial overlaps between clusters corresponding to different participants. In contrast, the clusters obtained from $\bm{r}_\mathrm{prop}$ are more distinct and clearly separated. This observation suggests that the proposed feature representation offers a potential advantage for identifying individuals.

\begin{figure}[tb]
	\centering
	\begin{minipage}{\linewidth}
		\centering
		\includegraphics[width=0.7\linewidth]{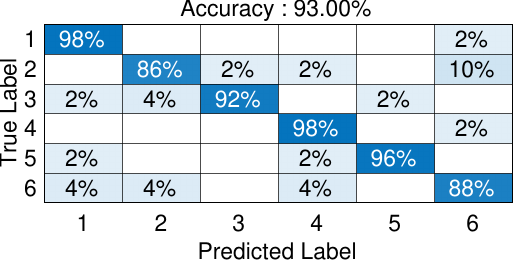}
		\subcaption{}
		\label{fig:confmat_amp}
	\end{minipage}
	\hspace{1mm}
	\begin{minipage}{\linewidth}
		\centering
		\includegraphics[width=0.7\linewidth]{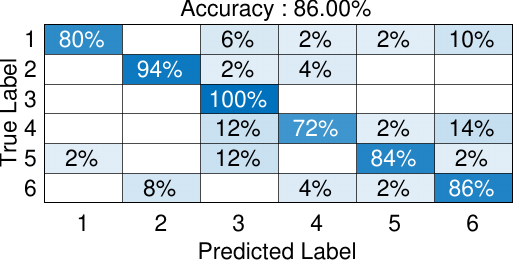}
		\subcaption{}
		\label{fig:confmat_ph}
	\end{minipage}
	\hspace{1mm}
	\begin{minipage}{\linewidth}
		\centering
		\includegraphics[width=0.7\linewidth]{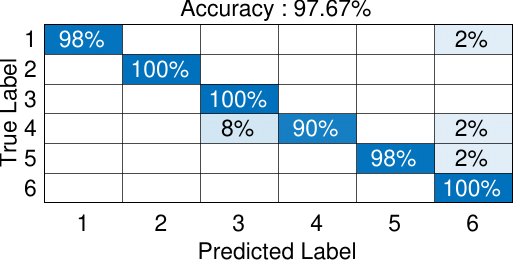}
		\subcaption{}
		\label{fig:confmat_prop}
	\end{minipage}
	\caption{Confusion matrices for (a) conventional method 1, (b) conventional method 2, and (c) proposed method with 60-s data samples.}
	\label{fig:confmat}
\end{figure}

Next, the feature vectors are used as inputs to a SVM classifier. 
The confusion matrices for the three methods are shown in Fig.~\ref{fig:confmat}. The identification accuracies obtained using $\bm{r}_\mathrm{amp}$, $\bm{r}_\mathrm{ph}$, and $\bm{r}_\mathrm{prop}$ were 93.00\%, 86.00\%, and 97.67\%, respectively. On average, the identification accuracy across all participants was 7.00 percentage points higher when using amplitude features $\bm{r}_\mathrm{amp}$ than when using phase features $\bm{r}_\mathrm{ph}$, suggesting that amplitude features are generally more effective for identifying individuals. However, the two feature types capture different characteristics. For example, for Participant~2, the identification accuracy was 86\% using amplitude features but 94\% when using phase features, corresponding to an 8 percentage point improvement. This result indicates that the effectiveness of each feature type varies across individuals, which is why improved performance was achieved by combining multiple feature representations.

\begin{table}[tb]
	\centering
	\caption{Accuracy and AUC Values for Each Method}
	\label{tab:acc_auc}
	\begin{tabular}{lcc}
		\toprule
		\textbf{Method}         & \textbf{Accuracy (\%)} & \textbf{AUC} \\
		\midrule
		Conv. 1                 & 93.00                  & 0.992        \\
		Conv. 2                 & 86.00                  & 0.969        \\
		Conv. 3~\cite{Kobayashi2024} & 93.33                  & 0.996        \\
		Prop.                   & 97.67                  & 0.999        \\
		\bottomrule
	\end{tabular}
\end{table}

Table~\ref{tab:acc_auc} summarizes the identification accuracy and area under the curve (AUC) values for each method. Notably, the proposed method, which combines $\bm{r}_\mathrm{amp}$, $\bm{r}_\mathrm{ph}$, and $\bm{r}_\mathrm{comp}$ for identification, demonstrated an improvement of 4.34 percentage points in accuracy and 0.003 in AUC compared with the conventional method~3, which relies solely on $\bm{r}_\mathrm{comp}$. These results indicate that the proposed method is effective for identifying individuals.

\begin{figure}[tb]
	\centering
	\includegraphics[width=0.5\linewidth]{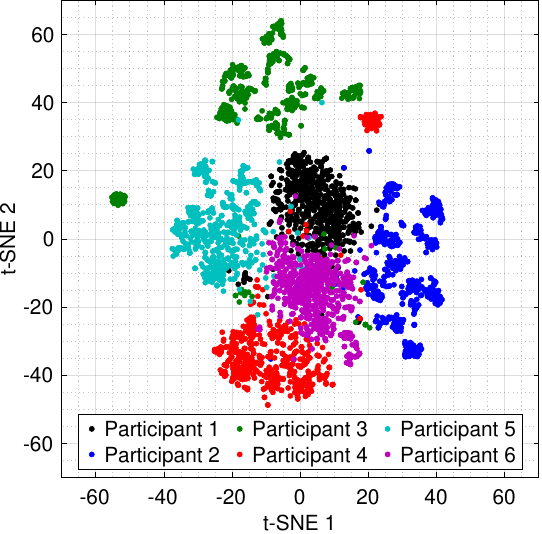}
	\caption{Two-dimensional t-SNE visualization of the $\bm{r}_\mathrm{prop}$ feature vectors for 5-s data samples.}
	\label{fig:tsne_prop_5s}
\end{figure}

\begin{figure}[tb!]
	\centering
	\includegraphics[width=0.75\linewidth]{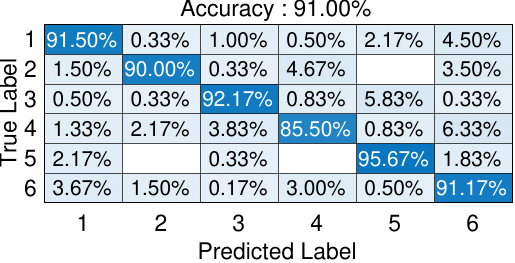}
	\caption{Confusion matrix for $\bm{r}_\mathrm{prop}$ with 5-s data samples.}
	\label{fig:confmat_prop_5s}
\end{figure}

Next, we evaluated the individual identification performance for shorter data samples with a length of 5~s. For this case, the evaluation was conducted for only the proposed method using $\bm{r}_\mathrm{prop}$.
We reported the individual identification performance of $\bm{r}_\mathrm{comp}$ with 5-s samples in previous work~\cite{Kobayashi2024}. Figure~\ref{fig:tsne_prop_5s} shows a two-dimensional t-SNE visualization of the $\bm{r}_\mathrm{prop}$ feature vectors for the 5-s samples. The confusion matrix obtained for the proposed method is shown in Fig.~\ref{fig:confmat_prop_5s}, and indicates an identification accuracy of 91.00\%, which is an improvement of 5.36 percentage points in average accuracy in comparison with the value of 85.64\% reported in~\cite{Kobayashi2024}.

These results demonstrate the effectiveness of the proposed method---which combines features extracted from amplitude, phase, and complex signals---even when using 5-s data samples. However, the identification accuracy achieved with 5-s data did not reach the level obtained with 60-s data.
In particular, the accuracy for Participant~4 was only 85.50\%, highlighting the challenges associated with applying the proposed method to shorter durations of data.

\section{Conclusion}
In this study, we proposed a radar-based method for identification of individuals that combines three types of features extracted from spectrograms of complex-valued radar-received signals, as well as their amplitude and phase. Although the number of participants in the testing was limited to six, the evaluation was conducted using data acquired on multiple days, including both morning and afternoon measurements, thereby incorporating temporal variability in the physiological signals. Experimental results demonstrated that the proposed method achieved an identification accuracy of 97.67\% with an AUC of 0.999 for 60-s data samples, indicating highly reliable identification of individuals when provided with sufficient observation time.
Furthermore, compared with conventional methods based solely on complex-signal spectrogram features, the proposed method achieved improvements in identification accuracy of 4.34 and 5.36 percentage points for data lengths of 60~s and 5~s, respectively. Although the accuracy obtained with 5-s data samples reached 91.00\%, further improvement is required to achieve robust performance with shorter duration observations.

\section*{Acknowledgment}
\addcontentsline{toc}{section}{Acknowledgment}
\scriptsize
This work was supported in part by the SECOM Science and Technology Foundation; in part by the Japan Science and Technology Agency (JST) under Grant JPMJMI22J2 and Grant JPMJMS2296; in part by JSPS KAKENHI under Grant 21H03427, Grant 23H01420, and Grant 23K26115; and in part by the New Energy and Industrial Technology Development Organization (NEDO). This study involved human subjects. All experimental procedures were approved by the Ethics Committee of the Graduate School of Engineering, Kyoto University (Approval no.~202223).
\normalsize
\balance

\end{document}